\documentstyle[aps,epsfig,prl]{revtex}
\begin{document}
\title{Magnetopolaron effects in light reflection and absorption\\
of a three-level system in a QW}
\author{I. G. Lang, L. I. Korovin}
\address{A. F. Ioffe Physico-Technical Institute, Russian
Academy of Sciences, 194021 St. Petersburg, Russia}
\author{D.A. Contreras-Solorio, S. T. Pavlov\cite{byline1}}
\address {Escuela de Fisica de la UAZ, Apartado Postal C-580,
98060 Zacatecas, Zac., Mexico}
\twocolumn[\hsize\textwidth\columnwidth\hsize\csname
@twocolumnfalse\endcsname
\date{\today}
\maketitle
\widetext
\begin{abstract}
\begin{center}
\parbox{6in}
{ The  light absorption ${\cal A}$ and reflection ${\cal R}$
coefficients for a three-level system in a quantum well (QW) in a
strong magnetic field {\bf H}, directed perpendicularly to the
QW's plane, have been calculated. The energy levels correspond to
the ground state and to magnetopolaron energy levels
$\hbar\omega_{1(2)}$ with the inverse lifetimes $\gamma_{r1(2)}$
and $\gamma_{1(2)}$. The inverse radiative lifetime of an EHP
$\gamma_{r\xi}(\bf{\cal K}_{\perp})\not=0$ only under condition
${\cal K}_{\perp}\le\omega_\xi n/c$, where $\hbar\omega_\xi$ is
the EHP energy. It has been shown that $\gamma_{r\xi}(\bf{\cal
K}_{\perp})$ is proportional to $H$. The values
$\gamma_{ra}(\bf{\cal K}_{\perp})$ and $\gamma_{rb}(\bf{\cal
K}_{\perp})$ for an excitation consisting of a hole and a usual
magnetopolaron have been calculated. The index $a(b)$ designates a
magnetopolaron upper (lower) term.  In the resonance at
$H_{res}=m_ec\omega_{LO}/|e|$~~
$\gamma_{ra}^{res}=\gamma_{rb}^{res}=\gamma_{r\xi_0}^{res}/2$,
where the index $\xi_0$ corresponds to the EHP with $n_e=n_h=1$.
$\gamma_{ra}$ and $\gamma_{rb}$ are strongly dependent on
$H-H_{res}$. A numerical estimate for GaAs is:
$\gamma_{r\xi_0}^{res}=\gamma_0=5.35\cdot10^{-5} eV.$
 The
dependencies ${\cal A}(H)$ and ${\cal R}(H)$ for various values of
$\omega_l$ in the magnetophonon resonance vicinity have been
obtained. The existence of "combined"and "weak" polarons has been
predicted. The resonant value $H_{res}$ for the combined polaron
case depends on a QW's depth and width.}
\end{center}
\end{abstract}

]
\narrowtext
\section{Introduction}

        When a polaron state is formed in a magnetic field
the role of the electron-phonon interaction grows sharply
under the  resonant condition
\begin{equation}
\label{1}
\omega_{LO}=j\Omega, ~~~~j=1, 2, 3,.....  ,
\end{equation}
$\omega_{LO}$ is the LO phonon frequency,
$\Omega$  is the electron (hole) cyclotron frequency.
The intersections of the electron-phonon system
energy levels as functions of magnetic field stem (see Fig.1).
A transition to the magnetopolaron states results in
energy levels splitting in crossing points. For the first time
magnetopolaron energy level splitting was discovered in light
absorption in bulk $InSb$.~\cite{1}

         Polaron state formation takes place in both three
(3D) and two dimensional (2D) semiconducting systems. These states
are very important in frequency dependence formation of
magnetooptical effects, such as reflection, interband absorption,
cyclotron resonance and Raman scattering (see, for instance, the
reviews Refs.~\onlinecite {2,3,4}). The systems distinguish by the
electron (hole) spectrum: in 3D systems these are  one-dimensional
Landau bands , whereas in  2D structures these are discrete energy
levels. This difference results in different splitting of the
energy levels of the electron-phonon system: In 3D systems it is
the value of order $\alpha^{2/3}\hbar\omega_{LO}$,~\cite{5}
whereas in 2D structures it is the value of order
$\alpha^{1/2}\hbar\omega_{LO}$,~\cite{6,7,8,9}  where $\alpha$ is
Fr\"ohlich's non-dimensional electron-LO phonon coupling constant.
~\cite{10}

        A perfect single QW of a finite depth is considered below
as an example of
a 2D system. The inhomogeneous broadening effects of the energy levels
are ignored. We shall calculate the non-dimensional reflection coefficient
 ${\cal R}$
and the non-dimensional light absorption coefficient ${\cal A}$
which are determined as the relation of the reflected or absorbed
light flux to the incident one. There are both the EHP discrete
energy levels and the continuum spectrum. The optical effects are
due to the resonance of the energies of incident light
$\hbar\omega_l$ and the EHP or magnetopolaron discrete energy
levels. Far from the  $H$ and $\omega_l$ values, corresponding to
the magnetopolaron resonance, we will consider the system as a
{\it two-level} one (the first level is the crystal ground state
energy and the second is the EHP energy
 level).
In the vicinity of the magnetopolaron resonance we consider the system
as a {\it three-level} one, including the ground state energy and
two polaron levels.

        Light absorption and reflection in quasi-2D systems
had been continually considered earlier. In calculations of
absorption usually the low order perturbation theory on the
light-electron system interaction (see, for instance,
Refs.~\onlinecite{11,12,13})  had been used. The result for the
light absorption coefficient contains the factor
$\Delta_{\gamma_{\rho}}(\hbar\omega_l-E_\rho)$, where $E_\rho$  is
the electron excitation energy measured from the ground state
energy,
\begin{equation}
\label{2} \Delta_\gamma (E)={1\over
\pi}{\hbar\gamma/2\over E^2+(\hbar\gamma/2)^2}
\end{equation}
is the function transiting into the Dirac $\delta$ - function
when $\gamma\rightarrow 0$,
 $\gamma_\rho$  is the non-radiative inverse lifetime on the
energy level  $\rho$. This result implicates some self-contradiction:
The value $\Delta_\gamma (E)\rightarrow\infty$ at
$E=0$ and $\gamma\rightarrow 0$, whereas the non-dimensional
absorption coefficient ${\cal A}$ cannot exceed the value of one.
This self-contradiction is lifted if one goes out the perturbation theory
limits on the light-electron coupling constant, i. e.
the contributions of all the orders on this coupling constant
are summarized.
This summarizing means taking into account the
sequence of all the
absorption and reradiation processes of the light quantum
$\hbar\omega_l$.
There appears a new concept of the radiation lifetime
$\gamma^{-1}_{r\rho}$ of an electronic excitation which
has been introduced firstly for the excitons in a QW at
$H=0$.~\cite{14}

        At first, the new approach had been used to describe
light reflection from QW ~\cite{12,14,15,16} at frequencies
$\omega_l$ close to an exciton energy, then an appropriate theory
for light absorption ~\cite{17,18} has been created. It has been
shown that the previous results obtained with the help of the
perturbation theory  are applicable under condition
\begin{equation}
\label{3}
\gamma_{r\rho}<<\gamma_\rho.
\end{equation}
        The new theory has to be used to describe the features
of light reflection and absorption by ideal QWs in magnetic fields
because the non-radiative values $\gamma_\rho$  for the discrete
energy levels of EHPs are small and the condition of Eq.
(3) may be unfeasible. We will calculate the coefficients
${\cal R}$ and ${\cal A}$ for the case of the incident light
perpendicular to the QW plane, taking into account the magnetopolaron
effects.

        The paper is organized as follows. The magnetopolaron classification
is described in Seq. II, the expressions for the electric
fields left and right from a QW at the normal irradiation
by the incident light in the case of the multilevel system are
obtained in Seq. III. The formulae for light reflection and
absorption in a QW are calculated in Seq. IY. In Seq. Y and Seq. YI,
the inverse radiative lifetime for an EHP in QW in magnetic
field and the inverse radiative lifetimes of two magnetopolaron
states are calculated, respectively. The inverse non-radiative
lifetimes of the magnetopolaron states are calculated in Seq. YII.
The numerical calculation results for light absorption and
reflection are given in Seq. YIII.

\section{Magnetopolaron classification}

        Fig. 1 shows (the full lines) the non-dimensional
electron-phonon system energy levels $E/\hbar\omega_{LO}$
in a QW, pertaining to the size-quantization quantum number $l$,
as functions of $\Omega/\omega_{LO}=j$,
\begin{equation}
\label{4}
\Omega=|e|H/m_{e(h)}c
\end{equation}
is the cyclotron frequency, $e$ is the electron charge, $m_{e(h)}$
is the electron (hole) effective mass. $E$ is measured from the
energy $\varepsilon_l^{e(h)}$ corresponding to the $l$
size-quantization energy level. It is supposed that all the
phonons relevant to the magnetopolaron creation have the same
non-dispersal frequency  $\omega_{LO}$.

        The polaron states correspond to the crossing points.
The "double" polarons are denoted with the black circles and
appropriate to crossings of two terms only. Let us consider
some crossing point corresponding to the number $j$ (see Eq. (1)).
$n$ is the Landau quantum number of the energy level coming through
this crossing point, the phonon number $N=0$ (see Fig. 1).
Then the condition of the double polaron existence is
\begin{equation}
\label{5}
2j>n\ge j.
\end{equation}

        It is easy to note that the only double polaron  $A$
corresponds to $j=1$. Two double polarons $D$ and $E$
correspond to $j=2$,
three double polarons $F$, $K$ and $L$ correspond to $j=3$
and so on. The polarons on the left of the value  $\Omega/\omega_{LO}=1/3$
are not noted in Fig.1. The triple polarons (corresponding
to  three terms crossings) are above the double polarons,
the quateron polarons are above the triple polarons and so on.
There are $j$ polarons of every sort at the given $j$.
For the first time the triple polarons in bulk semiconductors
and in QWs have been considered in
Ref.~\onlinecite{19} and in Refs.~\onlinecite{20,21,22}, respectively.

        Note that the Landau level's equidistancy is needed
for three and more levels crossings. The triple polaron theory~
\cite{22} is relevant to the case when the energy corrections due
to the band non-parabolicity or excitonic effects are smaller than
the triple polaron splitting. But in the case of the double
polarons, the equidistancy violation is not an obstacle because
crossing of two terms exists at any rate.

        All the above mentioned polarons correspond to the integer values of
$j$. At Fig. 1 there are some other crossings of the terms with
the size-quantization quantum number $l$ (solid lines) marked with
empty circles. They are relevant to the fractional values of $j$.
Because these crossings are characterized by the values $\Delta
N\ge 2$, the real direct transitions with emitting of {\it one} LO
phonon are impossible. We call such polarons {\it weak} ones. To
calculate splittings, one has to take into account the virtual
transitions between the crossing terms or two-phonon interaction.
Splittings of the weak polarons $\Delta E_w$ are much smaller than
for the polarons with integer $j$. The contributions of the
intermediate transitions into the value $\Delta E_w$ are of higher
order on $\alpha$ than $\alpha^{1/2}$.

        The crossing depiction becomes much more complicated when
two or more values of the size-quantization quantum numbers $l$
are taken into account.Besides the ordinary polarons corresponding
to the energy level $l^{\prime}$ (for example the polaron $A^{\prime}$)
 the "combined" polarons ~\cite{13} appear
where electron-phonon interaction bounds  two electron energy
levels with different Landau quantum numbers $n$ and different $l$
(or with different $l$ and the same $n$). Two terms corresponding
to the quantum number $l^{\prime}$ (dotted lines) and the combined
polaron $P$ and $Q$ positions (tilted crosses) are shown in Fig.
1. The combined weak polaron is marked with $R$.

        The combined polarons have an interesting feature: The
corresponding resonant magnetic field values depend on the distance
$\Delta\varepsilon=\varepsilon_{l^{\prime}}-\varepsilon_l$
between the size-quantization energy levels $l$ and $l^{\prime}$
and consequently on the QW's depth and width. With the help of Fig. 1
one can obtain easily
\begin{equation}
\label{6}
\left({\Omega\over\omega_{LO}}\right)_P=
1-{\Delta\varepsilon\over\hbar\omega_{LO}},~~
\left({\Omega\over\omega_{LO}}\right)_Q=
1+{\Delta\varepsilon\over\hbar\omega_{LO}}.
\end{equation}

        One more combined polaron sort ~\cite{13} exists under
condition
\begin{equation}
\label{7}
\Delta\varepsilon=\hbar\omega_{LO},
\end{equation}
when, for instance, the terms $l^{\prime}, n=0, N=0$ and $l, n=0, N=1$
coincide at any  magnetic field values. The definite separation
between the energy levels $l$ and $l^{\prime}$ is necessary
to satisfy the condition (7). This could be done by choosing
the QW's depth and width.

        The situation depicted in Fig. 1 is applicable when the energy
separations between the neighbor size-quantization levels $l$,
$l-1$, $l+1$ are much more than the values $\Delta E$ of polaron
splittings. So long as the separations diminish with growing of
the QW's width, some restriction is implicated on the value $d$
from above (see the numerical estimates in Ref.~\onlinecite{23}).

\section{Electric fields out of QW}
        Let us suppose that an electromagnetic (EM) wave (or some
superposition of EM waves, corresponding to a light impulse) falls
perpendicular on a single QW from the left. The electric field
intensity is \footnote{One can remove the complex conjugate term
and introduce (instead of ${\cal E}_{0\alpha}(\omega)$)
$$E_{0\alpha}(\omega)={\cal E}_{0\alpha}(\omega)+ {\cal
E}_{0\alpha}^*(-\omega)=2\pi
E_0[e_{l\alpha}D_0(\omega)+e_{l\alpha}^* D_0^*(-\omega)].$$}
\begin{equation}
\label{8}
E_{0\alpha}(z, t)={1\over 2\pi}
\int_{-\infty}^{\infty}
d\omega e^{-i\omega(t-zn/c)}{\cal E}_{0\alpha}(\omega)+c.c.,
\end{equation}
where $n$ is the barrier refraction index.

        Let us introduce the designations
\begin{equation}
\label{9}
{\cal E}_{0\alpha}(\omega)=2\pi E_0e_{l\alpha}
{\cal D}_0(\omega),
\end{equation}
where $E_0$ is the scalar amplitude, ${\bf e}_l$ is the
incident light polarization vector
; ${\cal D}_0(\omega)$ is a frequency function;
\begin{equation}
\label{10}
{\cal D}_0(\omega)=\delta(\omega-\omega_l)
\end{equation}
for monochromatic light of frequency $\omega_l$.

        It is supposed that the incident EM wave has a circular
polarization
\begin{equation}
\label{11}
{\bf e}_l={1\over\sqrt{2}}({\bf e}_x\pm i{\bf e}_y),
\end{equation}
where ${\bf e}_x, {\bf e}_y$ are the unit vectors along
the  $x$ and $y$ axis, respectively.

        The incident EM wave generates the excited electronic
states with the energies
characterized by the index $\rho$. For instance, in a
magnetic field the index $\rho$ for the EHP includes
$l_e, l_h, n_e=n_h$. For the infinitely deep QW $l_e=l_h$.
The state $\rho$ is characterized by the energy $\hbar\omega_\rho$,
which is measured from the ground state energy, by the inverse radiative
lifetime $\gamma_{r\rho}$ and by the inverse non-radiative
lifetime $\gamma_{\rho}$. Let us consider the QW with
$d<<c/\omega_ln$.

        Then the electric field intensities ${\bf E}_{left(right)}(z, t)$
on the left and right of the QW are determined as ~\cite{24}
\begin{equation}
\label{12}
{\bf E}_{left(right)}(z, t)={\bf E}_0(z, t)+
\Delta{\bf E}_{left(right)}(z, t),
\end{equation}
\begin{eqnarray}
\label{13} \Delta E_{\alpha~ left(right)}(z, t)&=&E_0e_{l\alpha}
\int_{-\infty}^{\infty} d\omega e^{-i\omega(t\pm zn/c)}{\cal
D}(\omega)+\nonumber\\ &+& c.c.,
\end{eqnarray}
where the upper (lower) sign is relevant to the index $left$
($right$). The frequency partitition is determined as \footnote
{It is supposed in Eq. (15) that both circular polarizations
corresponds to the  excitation of one of two types of EHPs
the energies of which are equal each other
(see Eq. (88) below).}

\begin{equation}
\label{14}
{\cal D}(\omega)=-{4\pi\chi(\omega){\cal D}_0(\omega)\over
1+4\pi\chi(\omega)},
\end{equation}
\begin{eqnarray}
\label{15}
\chi(\omega)&=&{i\over 4\pi}
\sum_\rho{\gamma_{r\rho}\over 2}
[(\omega-\omega_\rho+i\gamma_\rho/2)^{-1}+\nonumber\\
&+&(\omega+\omega_\rho+i\gamma_\rho/2)^{-1}].
\end{eqnarray}
Eqs. (12)--(15) determine the electric fields on the both
sides of the QW and, consequently, the values of the energy  fluxes
in transmitted and reflected light for the multilevel system.
The second term in the square brackets in the RHS of Eq. (15)
is non-resonant, but it is necessary to satisfy the relation
\begin{equation}
\label{16}
\chi^*(\omega)=\chi(-\omega).
\end{equation}
        In a real QW of a finite depth in magnetic field,
there are always discrete energy levels (one at least) together
with the continuous spectrum. Thus, the index $\rho$  in the RHS
of Eq. (15) takes both discrete and continuous values. Sometimes
in a magnetic field, the large level number has to be taken into
account (see ~\cite{25} where the results of an excitation with an
asymmetrical light impulse have been investigated). Discussing a
two-level system we mean a situation when all the energy levels
can be neglected, but one which is  resonant with $\hbar\omega_l$.

        The absorption and reflection coefficients for a two-level system
with one excitonic energy level at $H=0$ have been calculated in
Refs.~\onlinecite{14,15,16,17,18}. The electric field
intensities are expressed as
\begin{eqnarray}
\label{17}
E_{\alpha~ left}&=&E_0e_{l\alpha}
\left\{e^{-i\omega_l(t-zn/c)}-\right.\nonumber\\
&-&\left.{i\gamma_r\over 2
(\omega_l-\omega+i\Gamma/2)}e^{-i\omega_l(t+zn/c)}\right\}+c.c.,
\end{eqnarray}
\begin{eqnarray}
\label{18}
E_{\alpha~ right}&=&E_0e_{l\alpha}
e^{-i\omega_l(t-zn/c)}\times\nonumber\\
&\times&\left\{1-{i\gamma_r\over 2
(\omega_l-\omega+i\Gamma/2)}\right\}+c.c.,
\end{eqnarray}
\begin{equation}
\label{19}
\Gamma=\gamma+\gamma_r.
\end{equation}
One can obtain Eq. (17)  and Eq. (18) with the help of Eqs. (10),
(12)--(15) if one term in the sum on $\rho$ in the RHS
of Eq. (15) is preserved using the designations
\begin{equation}
\label{20}
\omega_\rho=\omega,~~~~~\gamma_{r\rho}=\gamma_r,~~~~~
\gamma_\rho=\gamma,
\end{equation}
and neglecting the second non-resonant term in the square brackets
in the RHS of Eq. (15).

        The three-level system consists of the ground state of an
electronic system and two  energy levels $\hbar\omega_{1(2)}$
with the inverse radiative $\gamma_{r1(2)}$ and non-radiative
$\gamma_{1(2)}$ lifetimes. The calculation of the electric field
intensities for the three-level system is a little more
complicated because one has to solve the square equation (see Eq. (23)
below). Applying Eqs. (10), (12)--(15) and omitting the non-resonant
term in the RHS of Eq. (15) one obtains
\begin{eqnarray}
\label{21}
&E&_{\alpha~ left}(z, t)=E_0e_{l\alpha}
\left\{e^{-i\omega_l(t-zn/c)}-ie^{-i\omega_l(t+zn/c)}\right.\times
\nonumber\\
&\times&
\left.\left[{\bar{\gamma}_{r1}\over 2(\omega_l-\Omega_1+iG_1/2)}
+{\bar{\gamma}_{r2}\over 2(\omega_l-\Omega_2+iG_2/2)}\right]\right\}
+\nonumber\\&+&c.c.,
\end{eqnarray}
\begin{eqnarray}
\label{22}
&E&_{\alpha~ right}(z, t)=E_0e_{l\alpha}e^{-i\omega_l(t-zn/c)}\times
\nonumber\\
&\times&\left\{1-i\left[{\bar{\gamma}_{r1}\over 2(\omega_l-\Omega_1+iG_1/2)}
+{\bar{\gamma}_{r2}\over 2(\omega_l-\Omega_2+iG_2/2)}\right]\right\}\nonumber\\
&+&c.c. .
\end{eqnarray}

The complex quantities $O_1=\Omega_1-iG_1/2$ and $O_2=\Omega_2-iG_2/2$~~~
($\Omega_{1(2)}$ and $G_{1(2)}$ are real on their definition)
are the precise solutions of the equation
\begin{eqnarray}
\label{23}
(O&-&\omega_1+i\gamma_1/2)(O-\omega_2+i\gamma_2/2)+\nonumber\\
&+&i{\gamma_{r1}\over 2}(O-\omega_2+i\gamma_2/2)+\nonumber\\
&+&i{\gamma_{r2}\over 2}(O-\omega_1+i\gamma_1/2)=0,
\end{eqnarray}
which are equal
\begin{equation}
\label{24}
(\Omega-iG/2)_{1,2}=
{1\over 2}\left[\tilde{\omega}_1+\tilde{\omega}_2\pm
\sqrt{(\tilde{\omega}_1-\tilde{\omega}_2)^2-\gamma_{r1}\gamma_{r2}}
\right],
\end{equation}
where
\begin{equation}
\label{25}
\tilde{\omega}_{1(2)}=\omega_{1(2)}-i\Gamma_{1(2)}/2,~~~
\Gamma_{1(2)}=\gamma_{1(2)}+\gamma_{r1(2)}.
\end{equation}
Note that the value in the square root in Eq. (24) is, generally
speaking, a complex one. In the RHS of Eq. (24) the upper (lower)
sign corresponds to the index 1(2). The following designations are used
\begin{equation}
\label{26}
\bar{\gamma}_{r1}=\gamma_{r1}+\Delta\gamma,~~~~~~
\bar{\gamma}_{r2}=\gamma_{r2}-\Delta\gamma,
\end{equation}
\begin{eqnarray}
\label{27}
\Delta\gamma&=&{\gamma_{r1}[\Omega_2-\omega_2-i(G_2-\gamma_2)/2]
\over\Omega_1-\Omega_2+i(G_2-G_1)/2}+\nonumber\\
&+&{\gamma_{r2}[\Omega_1-\omega_1-i(G_1-\gamma_1)/2]\over
\Omega_1-\Omega_2+i(G_2-G_1)/2}.
\end{eqnarray}
Comparing Eq. (21) and Eq. (22) with  Eq. (17) and Eq. (18)
for the two-level system, one can see that the addition of
the contributions from the energy levels 1 and 2 is accompanied
by the substitutions:
$\Omega_{1(2)}$ instead of $\omega_{1(2)}$, $G_{1(2)}$  instead of
$\Gamma_{1(2)}$ and
$\bar{\gamma}_{{r1(2)}}$ instead of $\gamma_{{r1(2)}}$.

        To calculate the electric field intensities for the
four-level system one has to solve a third order equation
and so on. It is impossible to solve this task precisely for
an arbitrary number $\rho$ of the energy levels. However, in the case
\begin{equation}
\label{28}
\gamma_{r\rho}<<\gamma_\rho,
\end{equation}
when the  perturbation theory on light-electron system interaction
is applicable (what results in the possibility of neglecting the
term $4\pi\chi(\omega)$
in the denominator in the RHS of Eq. (14)), we obtain the result
for the arbitrary number  of the energy levels
\begin{eqnarray}
\label{29}
&E&_{\alpha~ left}^{many}(z, t)=E_0e_{l\alpha}
\left\{e^{-i\omega_l(t-zn/c)}-\right.\nonumber\\
&-&\left.ie^{-i\omega_l(t+zn/c)}\sum_\rho
{\gamma_{r\rho}\over 2(\omega_l-\omega_\rho+i\gamma_\rho/2)}\right\}
+c.c.,
\end{eqnarray}
\begin{eqnarray}
\label{30}
&E&^{many}_{\alpha~ right}(z, t)=E_0e_{l\alpha}e^{-i\omega_l
(t-zn/c)}\times\nonumber\\
&\times&\left[1-i\sum_\rho{\gamma_{r\rho}\over 2(\omega_l-\omega_\rho
+i\gamma_\rho/2)}\right]+c.c.,
\end{eqnarray}
\section{Light reflection and absorption coefficients.}

        Having the electric field intensities
out of the QW we can calculate the reflection and absorption coefficients of
light. The Umov-Poynting vector ${\bf S}_{left}$ on the left of the QW
 is
\begin{equation}
\label{31}
{\bf S}_{left}={\bf S}_0+\Delta{\bf S}_{left},
\end{equation}
where
\begin{equation}
\label{32}
{\bf S}_0={c\over 2\pi}E_0^2{\bf e_z}
\end{equation}
is the incident light flux,  $\Delta{\bf S}_{left}$ is the
reflected light flux directed along $~~-{\bf e}_z$.
        The light reflection coefficient is determined as
\begin{equation}
\label{33}
{\cal R}={|\Delta{\bf S}_{left}|\over|{\bf S}_0|},
\end{equation}
the non-dimensional light absorption coefficient is determined as
\begin{equation}
\label{34}
{\cal A}={|{\bf S}_{left}-{\bf S}_{right}|\over|{\bf S}_0|},
\end{equation}
and the light transition coefficient is
\begin{equation}
\label{35}
{\cal T}=1-{\cal R}-{\cal A}={|{\bf S}_{right}|\over|{\bf S}_0|}.
\end{equation}
With the help of Eq. (17) and Eq. (18) we obtain the results for the
two-level system
\begin{equation}
\label{36}
{\cal R}={(\gamma_r/2)^2\over (\omega_l-\omega)^2+(\Gamma/2)^2},
\end{equation}
\begin{equation}
\label{37}
{\cal A}={\gamma\gamma_r/2\over (\omega_l-\omega)^2+(\Gamma/2)^2},
\end{equation}
obtained earlier in Refs.~\onlinecite{14,15,16,17,18}.
Eq. (37) contains the very important result: Light absorption
in a QW is absent completely when $\gamma=0$.~\cite{17,18}

        Let us consider two limit cases:  $\gamma_r<<\gamma$ and
$\gamma_r>>\gamma$. In the case  $\gamma_r<<\gamma$ the perturbation
theory on light-electron interaction is applicable in the lowest order.
In this approximation, the values of
the light absorption and reflection coefficients are of the
second and fourth order on interaction, respectively.
From Eq. (36) and Eq. (37) we obtain

\begin{equation}
\label{38}
{\cal R}\simeq{\pi\hbar\over 2}{\gamma_r^2\over\gamma}\Delta_\gamma
[\hbar(\omega_l-\omega)],
\end{equation}
\begin{equation}
\label{39}
{\cal A}\simeq\pi\hbar\gamma_r\Delta_\gamma[\hbar(\omega_l-\omega)],
\end{equation}
where $\Delta_\gamma(E)$ is determined in Eq. (2).

        At $\gamma_r<<\gamma$
\begin{equation}
\label{40}
{\cal A}<<1,~~~{\cal R}<<{\cal A}.
\end{equation}
        In the opposite case $\gamma_r>>\gamma$ in Eq. (36) and Eq. (37)
$\Gamma$ has to be substituted approximately by $\gamma_r$.
Then we obtain that in the resonance ${\cal R}(\omega_l=\omega)=1$,
which means the total light reflection, ${\cal A}<<1$.

        Thus, light absorption is small for both cases
$\gamma_r<<\gamma$ and $\gamma_r>>\gamma$.  The maximum value
${\cal A}(\omega_l=\omega)=1/2$ is reached in the case $\gamma_r=\gamma$.

        With the help of Eq. (33), Eq. (34) and Eq. (21), Eq. (22)
we obtain for the three-level system
\begin{eqnarray}
\label{41}
{\cal R}&=&{1\over 4Z}\{[\gamma_{r1}(\omega_l-\omega_2)+
\gamma_{r2}(\omega_l-\omega_1)]^2+\nonumber\\
&+&(\gamma_{r1}\gamma_2+\gamma_{r2}\gamma_1)^2/4\},
\end{eqnarray}
\begin{eqnarray}
\label{42}
{\cal A}&=&{1\over 2Z}\{\gamma_{r1}\gamma_1[(\omega_l-\omega_2)^2+
(\gamma_2/2)^2]+\nonumber\\
&+&\gamma_{r2}\gamma_2[(\omega_l-\omega_1)^2+
(\gamma_1/2)^2]\},
\end{eqnarray}
where
\begin{equation}
\label{43}
Z=[(\omega_l-\Omega_1)^2+(G_1/2)^2][(\omega_l-\Omega_2)^2+(G_2/2)^2].
\end{equation}
It follows from Eq. (42) that light absorption by the three-level system
is equal 0, if
\begin{equation}
\label{44}
\gamma_1=\gamma_2=0.
\end{equation}
Applying the fact that the quantities $(\Omega-iG/2)_{1,2}$ are
the roots of Eq. (23), we transform the denominator in Eq. (43) to
the form
\begin{eqnarray}
\label{45}
Z&=&[(\omega_l-\omega_1)(\omega_l-\omega_2)
-(\gamma_{r1}\gamma_2+\gamma_{r2}\gamma_1
+\gamma_1\gamma_2)/4]^2+\nonumber\\
&+&[(\omega_l-\omega_1)\Gamma_2+(\omega_l-\omega_2)\Gamma_1]^2/4.
\end{eqnarray}

        Let us obtain the simplified expressions for
${\cal R}$ and ${\cal A}$ for the different limits. It is
convenient to use Eq. (41) and Eq. (42) with the substitution Eq.
(43) or, sometimes, with Eq. (45). The first limit case
corresponds to the inequalities
\begin{equation}
\label{46}
\gamma_{r1(2)}<<\gamma_{1(2)},
\end{equation}
when the perturbation theory on light-electron interaction is
applicable what corresponds to neglecting by the term $4\pi\chi(\omega)$
in the RHS of Eq. (14). Under condition Eq. (46) in the RHS
of Eq. (43) we suppose
$\Omega_{1(2})\simeq\omega_{1(2)},G_{1(2)}=\gamma_{1(2)}$
and obtain
\begin{eqnarray}
\label{47}
{\cal R}&\simeq&{(\gamma_{r1}/2)^2\over(\omega_l-\omega_1)^2+(\gamma_1/2)^2}
+{(\gamma_{r2}/2)^2\over(\omega_l-\omega_2)^2+(\gamma_2/2)^2}+\nonumber\\
&+&{\gamma_{r1}\gamma_{r2}\over 2}\times\nonumber\\
&\times&{(\omega_l-\omega_1)(\omega_l-\omega_2)+\gamma_{1}\gamma_{2}/4\over
[(\omega_l-\omega_1)^2+(\gamma_1/2)^2]
[(\omega_l-\omega_2)^2+(\gamma_2/2)^2]},
\end{eqnarray}
\begin{equation}
\label{48}
{\cal A}\simeq
{\gamma_{r1}\gamma_1/2\over
(\omega_l-\omega_1)^2+(\gamma_1/2)^2}
+{\gamma_{r2}\gamma_2/2\over
(\omega_l-\omega_2)^2+(\gamma_2/2)^2}.
\end{equation}
According to Eq. (48) the absorption coefficient ${\cal A}$ is the
sum of contributions Eq. (39) from the levels 1 and 2, because
under condition Eq. (46) absorption is linear on the constants
$\gamma_{r1}$ and $\gamma_{r2}$. According to Eq. (47) the
reflection coefficient ${\cal R}$ is square-law on $\gamma_{r1}$
and $\gamma_{r2}$. Therefore it contains an interferential
contribution besides the separate levels contributions. The
functions ${\cal A}(\omega_l)$ and ${\cal R}(\omega_l)$ are
represented on Fig. 2 under condition Eq. (46) for the case
$\gamma_{r1}=\gamma_{r2}$, $\gamma_1=\gamma_2$.

        Generalizing Eq. (48) for the case of the arbitrary number of the
energy levels and under condition Eq. (28) one obtains
\begin{equation}
\label{49}
{\cal A}^{many}\simeq{1\over 2}\sum_\rho
{\gamma_{r\rho}\gamma_\rho
\over(\omega_l-\omega_\rho)^2+(\gamma_\rho/2)^2}.
\end{equation}
        The next limit case (opposite to the preceding one) is determined
by the conditions
\begin{equation}
\label{50}
\gamma_{r1(2)}>>\gamma_{1(2)}.
\end{equation}
We suppose
$\gamma_1=\gamma_2=0$ in the RHS of Eq. (41) and Eq. (42). Then ${\cal A}=0$
and one obtains
\begin{eqnarray}
\label{51}
&{\cal R}&(\gamma_1=\gamma_2=0)=
[(\gamma_{r1}+\gamma_{r2})/2]^2
(\omega_l-\omega_0)^2\times\nonumber\\
&\times&
[(\omega_l-\omega_1)^2(\omega_l-\omega_2)^2+\nonumber\\
&+&[(\gamma_{r1}+\gamma_{r2})/2]^2
(\omega_l-\omega_0)^2]^{-1},
\end{eqnarray}
where
\begin{equation}
\label{52}
\omega_0=
{\omega_1\gamma_{r2}+\omega_2\gamma_{r1}
\over\gamma_{r1}+\gamma_{r2}}.
\end{equation}
The peculiar properties of light reflection from the three-level
system follow from Eq. (51) in the case  of the dominant inverse
radiative lifetimes.
{\it At any quantities}
$\gamma_{r1}$ {\it and} $\gamma_{r2}$
{\it in the point}
$\omega_l=\omega_0$
{\it light reflection equals 0; in the points}
$\omega_l=\omega_1$ {\it and} $\omega_l=\omega_2$
{\it the reflection coefficient equals 1}.
Let us suppose $\gamma_{r1}=\gamma_{r2}=\gamma_r$. Then ${\cal R}=0$ at
$\omega_l=(\omega_1+\omega_2)/2$. At $\gamma_r>>(\omega_1-\omega_2)$
in the point $\omega_l=(\omega_1+\omega_2)/2$
there is a narrow minimum
descending to ${\cal R}=0$.
The half-width of this minimum on the half-depth equals to
$(\omega_1-\omega_2)^2/2^{3/2}\gamma_r$ .

        The functions ${\cal R}(\omega_l)$ are represented in Fig.
 3b under conditions
\begin{equation}
\label{53}
\gamma_{1}=\gamma_{2}=\gamma,
~~~~~\gamma_{r1}=\gamma_{r2}=\gamma_{r},~~~\gamma<<\gamma_{r}.
\end{equation}
The functions ${\cal A}(\omega_l)$ are represented on Fig. 3a
under the same condition Eq. (53).
Note a peculiar turn of  ${\cal A}(\omega_l)$ curves at
a fixed quantity $\gamma$   and at growing $\gamma_r$.
Transiting from the curve {\it 1} to the curve {\it 6} the quantity
$\hbar\gamma_r$ takes a successive row of values:
0.002, 0.005, 0.008,  0.04,  0.125,  0.5.
There are two maxima on the first curve at
$\hbar\gamma_r=0.002$. On the consecutive curves there is
one maximum which reaches its largest value
${\cal A}_{max}=0.5$ on the curve  {\it 5} ($\hbar\gamma_r=0.125$),
and afterwards its height ${\cal A}_{max}$  diminishes.
The quantity  ${\cal A}_0$ corresponding to the central point
$\omega_l=(\omega_1+\omega_2)/2$ and coinciding with the height
${\cal A}_{max}$ on the curves {\it 2} - {\it 6} is described by the
precise formula
\begin{equation}
\label{54}
{\cal A}_0=
{4\gamma_r\gamma[(\omega_1-\omega_2)^2+\gamma^2]
\over
[(\omega_1-\omega_2)^2
+2\gamma_r\gamma+\gamma^2]^2},
\end{equation}
which follows from Eq. (42) at $\omega_l=(\omega_1+\omega_2)/2$.
The quantity  ${\cal A}_0$ maximizes at
\begin{equation}
\label{55}
\gamma_{r0}={(\omega_1-\omega_2)^2+\gamma^2\over2\gamma}.
\end{equation}
Substituting the values  $\hbar(\omega_1-\omega_2)=0.005$
and $\hbar\gamma=0.0001$ in Eq. (55), what corresponds Fig. 3a,
one obtains $\hbar\gamma_{r0}=0.125$ and ${\cal A}_{max}=
{\cal A}_0=0.5$   (the curve {\it 5} on Fig.3a).

        The next limit realizes when the reverse lifetimes
$\gamma_{1(2)}$ and $\gamma_{r1(2)}$ are small in comparison with
the interlevel separation $\omega_1-\omega_2$. Let us suppose
\begin{equation}
\label{56}
\gamma_{1(2)}<<\omega_1-\omega_2, ~~~~~\gamma_{r1(2)}<<\omega_1-\omega_2.
\end{equation}
The interrelation between the non-radiative $\gamma_{1(2)}$ and
radiative $\gamma_{r1(2)}$ lifetimes can be an arbitrary one.
Then, if the frequency $\omega_l$  is close to
the resonance with one of the levels
, let us say $\omega_l\simeq\omega_1$,
one obtains from Eq. (41) and Eq. (42)
\begin{eqnarray}
\label{57}
{\cal R}&\simeq&
{\gamma_{r1}^2/4\over(\omega_l-\omega_1)^2+(\Gamma_1/2)^2},\nonumber\\
{\cal A}&\simeq&
{\gamma_{r1}\gamma_1/2\over(\omega_l-\omega_1)^2+(\Gamma_1/2)^2},
\end{eqnarray}
which coincides with the results of Eq. (36) and Eq. (37) for the
two-level system.

        Lastly we consider the case of the merging curves.
Supposing
\begin{equation}
\label{58}
\gamma_1=\gamma_2=\gamma,~~~\omega_1=\omega_2=\omega,
~~~\gamma_{r1}=\gamma_{r2}=\gamma_r,
\end{equation}
we obtain from Eq. (41) and Eq. (42)
\begin{equation}
\label{59}
{\cal R}\simeq
{\gamma_r^2\over(\omega_l-\omega)^2+(\gamma_r+\gamma/2)^2},
\end{equation}
\begin{eqnarray}
\label{60}
{\cal A}\simeq
{\gamma_r\gamma\over(\omega_l-\omega)^2+(\gamma_r+\gamma/2)^2}.
\end{eqnarray}
These results are similar to Eq. (36) and Eq. (37) for the two-level
system, but {\it\underline{the quantity}} $\gamma_r$
{\it\underline{is reduplicated}}. That means that
{\it in the case of a double degenerated excited level
the formulae for the two-level system with
\underline{the reduplicated value}
$\gamma_r$} {\it are applicable}.

\section{EHP radiative lifetime for QW in magnetic field.}

        One could calculate the EHP inverse radiative lifetime in
magnetic field indirectly: calculating the absorption coefficient
${\cal A}$ with the help of the perturbation theory and then,
comparing the obtained result with Eq. (39), to calculate $\gamma_r$.
The second way is the direct calculation of $\chi(\omega)$ (see Eq. (15)).
But we shall demonstrate the third way: The direct calculation
of $\gamma_r$ with the help of the Fermi Golden Rule
\begin{eqnarray}
\label{61}
\gamma_{r\eta}={2\pi\over\hbar}\sum_s|\langle s|U|\eta\rangle|^2
\delta(\hbar\omega_s-E_\eta).
\end{eqnarray}
We ascertained that all three calculation methods led to the same results
for $\gamma_r$.

        In Eq. (61) $|\eta\rangle, |s\rangle$ are the secondary quantized
wave functions.
The initial state wave function
$|\eta\rangle$ describes the EHP with the index set
$\eta$; the final state $|s\rangle$ describes the photon with the index set
$s$.  The interaction $U$ is written as
\begin{equation}
\label{62}
U=-{1\over c}\int d{\bf r}{\bf A}({\bf r}){\bf j}({\bf r}),
\end{equation}
\begin{equation}
\label{63}
{\bf A}({\bf r})=\left({2\pi\hbar\over V_0}\right)^{1/2}{c\over n}
\sum_s\omega_s^{-1/2}
(c_s{\bf e}_se^{i{\bf \kappa r}}+
c_s^+{\bf e}_s^*e^{-i{\kappa r}}).
\end{equation}
The index set $s$ involves the wave vector ${\bf\kappa}$ and the
polarization index $i$ (accepting two values), $c_s^+(c_s)$ is the
photon creation (annihilation) operator
, ${\bf e}_s$ and
$\hbar\omega_s$ are the polarization vector and energy, respectively;
$V_0$ is the normalization volume. The charge current density operator
is given by
\begin{equation}
\label{64}
{\bf j}({\bf r})={e\over m_0}
\sum_\xi
[{\bf p}_{cv}F_\xi^*({\bf r})a^+_\xi+
{\bf p}_{cv}^*F_\xi({\bf r})a_\xi],
\end{equation}
where $m_0$ is the bare electron mass, $a^+_\xi (a_\xi)$
is the EHP creation (annihilation) operator, $\xi$ is the index set,
\begin{equation}
\label{65}
F_\xi({\bf r})=\Psi_\xi({\bf r}_e={\bf r}_h={\bf r}),
\end{equation}
$\Psi_\xi({\bf r}_e,{\bf r}_h)$ is the smooth part
of the EHP wave function (in the effective mass approximation),
depending on the radius vectors  ${\bf r}_e,{\bf r}_h$
of the electron and hole, respectively,
 ${\bf p}_{cv}$ is the interband matrix element
of the momentum operator.
Let us stress that the index set  $\xi$ includes the indexes
$c$ and $v$ of the conductivity and valence bands, which can be
degenerate ones.

        Let us determine the wave functions $\Psi_\xi({\bf
r}_e,{\bf r}_h)$. That is well known that the electron (hole) wave functions,
corresponding to the gauge
${\bf A}={\bf A}(0, xH, 0)$ of the vector-potential, are
\begin{equation}
\label{66}
\psi_{c(v), n, k_y, l}^{e(h)}({\bf r})=\phi_n(x\pm a_H^2k_y)
{1\over\sqrt{L_y}}e^{ik_yy}\varphi_{c(v)l}(z).
\end{equation}
The upper sign "+" corresponds to the electron,
$a_H=\sqrt{{c\hbar/|e|H}}$, $L_y$ is the normalization length,
\begin{eqnarray}
\label{67}
\phi_n(x)={1\over\sqrt{a_H}}f_n(x/a_H),\nonumber\\
f_n(t)={1\over\sqrt{\pi^{1/2}2^nn!}}e^{-{t^2/2}}H_n(t),
\end{eqnarray}
$H_n(t)$ is the Hermitian polynomial.
The real functions $\varphi_{c(v)l}(z)$,
corresponding to the size-quantization quantum number $l$
for the QW of the finite depth, can be found, for instance, in
Ref.~\onlinecite{26}.

        At first we introduce an EHP wave function set consisting of
a product of the wave functions Eq. (66) for electrons and holes and
characterized by the index set
\begin{equation}
\label{68}
\zeta\rightarrow c,v, n_e, n_h, k_{ey}, k_{hy}, l_e, l_h.
\end{equation}
\begin{equation}
\label{69}
\Psi_\zeta({\bf r}_e, {\bf r}_h)=
\psi_{c, n_e, k_{ey}, l_e}^e
({\bf r}_e)
\psi_{v, n_h, k_{hy}, l_h}^h
({\bf r}_h).
\end{equation}
However one cannot calculate the EHP radiative lifetime with the help
of Eq. (69). One has to introduce the wave functions for a state of the EHP
created by light from the ground state of the system in a QW.
Therefore we go to the wave functions
\begin{eqnarray}
\label{70}
\Psi_\xi({\bf r}_e, {\bf r}_h)&=&
\sqrt{{2\pi a_H^2\over L_xL_y}}
\sum_{k_{ey}, k_{hy}}\delta_{K_y, k_{ey}+ k_{hy}}
e^{-ia_H^2K_xk_y}\times\nonumber\\
&\times&\Psi_\zeta({\bf r}_e, {\bf r}_h),
\end{eqnarray}
where
$$k_y={k_{ey}m_h- k_{hy}m_e\over M},~~~~M=m_e+m_h,$$
$m_e ( m_h)$ is the electron (hole) effective mass.

The functions of type of Eq. (70) have been introduced in
Ref.~\onlinecite{27},
see also Refs.~\onlinecite{28,29,30}. The index set is
\begin{equation}
\label{71}
\xi\rightarrow c,v, n_e, n_h, {\bf K}_{\perp}, l_e, l_h.
\end{equation}
It has been shown in Ref.~\onlinecite{29} that the functions Eq. (70)
are the eigenfunctions of the EHP momentum operator
$\hat{\bf P}_\perp$ in the $xy$ plane, corresponding to the eigenvalues
$\hbar {\bf K}_\perp$. The operator $\hat{\bf P}_\perp$ is determined
in Ref.~\onlinecite{31}.

        With the help of the coordinates
$${\bf r}_\perp={\bf r}_{e_\perp}-{\bf r}_{h\perp},~~~~~~
{\bf R}_\perp={m_h{\bf r}_{h\perp}+
m_e{\bf r}_{e\perp}
\over M}$$
of the relative and whole movement of the electron and hole in
the $xy$ plane
and summarizing on $k_{ey}, k_{hy}$ in the RHS of Eq. (70),
one obtains ~\cite{28,29,30}
\begin{eqnarray}
\label{72}
&\Psi&_\xi({\bf r}_e, {\bf r}_h)=
{1\over\sqrt{2\pi a_H^2L_xL_y}}\times\nonumber\\
&\times&\exp\left[i(K_xX+K_yY)-i{yX\over a^2_H}\right]\times\nonumber\\
&\times&\exp\left[i{m_e-m_h\over 2M}
(-{\bf K}_\perp{\bf r}_\perp+a_H^2K_xK_y)+{xy\over a^2_H}\right]
\times\nonumber\\
&\times&{\cal{K}}_{n_e, n_h}
\left({{\bf r}_\perp- {\bf r}_{\perp 0}\over a_H}\right)
\varphi_{cl_e}(z_e)\varphi_{vl_h}(z_h),
\end{eqnarray}
where
$${\bf r}_{\perp 0}=a_H^2{{\bf H}\times{\bf K}_\perp\over H}.$$

        The function ${\cal K}_{n, m}({\bf p})$
is determined by ~\cite{28}
\begin{eqnarray}
\label{73}
{\cal K}_{n, m}({\bf p})&=&\left[{min(n!, m!)\over max(n!, m!)}\right]^{1/2}
i^{|n-m|}e^{-p^2/4}\times\nonumber\\
&\times&\left({p\over\sqrt{2}}\right)^{|n-m|}
\exp[i(\phi -\pi/2)(n-m)]\times\nonumber\\
&\times&L_{min(n, m)}^{|n-m|}(p^2/2),
\end{eqnarray}
where
$$p=\sqrt{p_x^2+p_y^2},~~~~~~~\phi=\arctan(p_y/p_x),$$
$L_m^n(x)$ is the Laguerre polynomial.
The energy eigenvalues of the states Eq. (72)
are
\begin{eqnarray}
\label{74}
E_\xi=\hbar\omega_\xi&=&E_g+\varepsilon^e_{le}+\varepsilon^h_{lh}+\nonumber\\
&+&\hbar\Omega_e(n_e+1/2)+\hbar\Omega_h(n_h+1/2),
\end{eqnarray}
where  $E_g$ is the energy gap.

        With the help of Eq. (72) one obtains
\begin{eqnarray}
\label{75}
F_\xi({\bf r})&=&{1\over \sqrt{2\pi a^2_HL_xL_y}}\exp\left(
i{m_e-m_h\over 2M}a_H^2K_xK_y\right)\times\nonumber\\
&\times&{\cal K}_{n_e, n_h}\left(-{{\bf r}_{\perp 0}\over a_H}\right)
e^{i{\bf K}_\perp{\bf r}_\perp}\varphi_{cl_e}(z)\varphi_{vl_h}(z).
\end{eqnarray}
Let us calculate  $\gamma_{r\xi}$ for the index set $\xi$.
With the help of Eqs. (61) - (64) one obtains
\begin{eqnarray}
\label{76}
\gamma_{r\xi}&=&{(2\pi e)^2\over V_0(nm_0)^2}\sum_s
{|{\bf e}_s{\bf p}_{cv}|^2\over\omega_s}
\left|\int d{\bf r}e^{-i{\bf\kappa}{\bf r}}F_\xi({\bf r})
\right|^2\times\nonumber\\
&\times&\delta(\hbar\omega_s-E_\xi).
\end{eqnarray}
The substitution of Eq. (75) in  Eq. (76) results in
\begin{eqnarray}
\label{77}
\gamma_{r\xi}&=&\left({e\over a_Hnm_0}\right)^2
{1\over\hbar\omega_\xi}B_{n_e, n_h}(K_\perp){2\pi\over L_z}
\sum_{i, {\bf\kappa}}\delta_{{\bf\kappa}_\perp, {\bf K}_\perp}
\times\nonumber\\
&\times&|{\bf e}_s{\bf p}_{cv}|^2
|R_{l_e, l_h}(\kappa_z)|^2
\delta(\omega_s-\omega_\xi),
\end{eqnarray}
where
\begin{eqnarray}
\label{78}
B_{n, m}(K_\perp)&=&
\left|{\cal K}_{n, m}\left(-{{\bf r}_{\perp 0}\over a_H}\right)
\right|^2=
{min(n!, m!)\over max(n!, m!)}\times\nonumber\\
&\times&\exp\left(-{a_H^2K_\perp^2\over 2}\right)
\left({a_H^2K_\perp^2\over 2}\right)^{|n-m|}\times\nonumber\\
&\times&
\left[L_{min(n, m)}^{|n-m|}\left({a_H^2K_\perp^2\over 2}\right)\right]^2,
\end{eqnarray}
\begin{equation}
\label{79}
R_{l_e, l_h}(k)=\int_{-\infty}^\infty dz
e^{-ikz}\varphi_{cl_e}(z)\varphi_{vl_h}(z).
\end{equation}
Summing up in the RHS of Eq. (77) on ${\bf\kappa}_\perp$ and taking into
account that
$\omega_s=c\kappa/n,~~\kappa=\sqrt{\kappa^2_\perp+\kappa_z^2}$,
one obtains
\begin{eqnarray}
\label{80}
\gamma_{r\xi}&=&{2\pi e^2\over L_zm_0^2\hbar\omega_\xi a_H^2cn}
\times\nonumber\\
&\times&B_{n_e, n_h}(K_\perp)
\left|R_{l_e, l_h}(\sqrt{(\omega_\xi n/c)^2-K^2_\perp})\right|^2
\times\nonumber\\
&\times&\sum_{i, \kappa_z}
|{\bf e}_s{\bf p}_{cv}|^2
\delta\left(\sqrt{K^2_\perp+\kappa_z^2}-\omega_\xi n/c\right).
\end{eqnarray}
Summing up on $\kappa_z$ at $K_\perp <\omega_\xi n/c$, one obtains
\begin{eqnarray}
\label{81}
\gamma_{r\xi}&=&{e^2\Omega_0\over \hbar^2c^2m_0}
B_{n_e, n_h}(K_\perp)\times\nonumber\\
&\times&\left|R_{l_e, l_h}(\sqrt{(\omega_\xi n/c)^2-K^2_\perp})\right|^2
{1\over\sqrt{(\omega_\xi n/c)^2-K^2_\perp}}\times\nonumber\\
&\times&\sum_i
[|{\bf e}_i({\bf\kappa}^+){\bf p}_{cv}|^2+
|{\bf e}_i({\bf\kappa}^-){\bf p}_{cv}|^2];\nonumber\\
\gamma_{r\xi}&=&0,~~~~~~при~~~~~ K_\perp>\omega_\xi n/c,
\end{eqnarray}
where $\Omega_0=|e|H/m_0c$,
 ${\bf\kappa}^+({\bf\kappa}^-)$ is the photon wave vector
with the component ${\bf K}_\perp$ in the $xy$ plane
 and with the z-component equal to $\pm\sqrt{(\omega_\xi
n/c)^2-K^2_\perp}$.
Thus the EHP with the energy
$\hbar\omega_\xi$ and the wave vector ${\bf K}_\perp$
in the $xy$ plane emits the photons with the energies
$\hbar\omega_\xi$ and wave vectors  ${\bf\kappa}^+$
and ${\bf\kappa}^-$ if  $K_\perp<\omega_\xi n/c$.

        Let us direct the $x$ axis along  ${\bf K}_\perp$.
Summarizing on the photon polarizations at $K_\perp<\omega_\xi n/c$
one obtains
\begin{eqnarray}
\label{82}
\gamma_{r\xi}&=&{2e^2\Omega_0\over\hbar^2cnm_0\omega_\xi}
B_{n_e, n_h}(K_\perp)
\times\nonumber\\
&\times&\left|R_{l_e, l_h}(\sqrt{(\omega_\xi n/c)^2-K^2_\perp})\right|^2
\times\nonumber\\
&\times&\left[{\sqrt{(\omega_\xi n/c)^2-K^2_\perp}\over\omega_\xi n/c}
|p_{cvx}|^2+\right.\nonumber\\
&+&{\omega_\xi n/c\over\sqrt{(\omega_\xi n/c)^2-K^2_\perp}}|p_{cvy}|^2
+\nonumber\\
&+&\left.{K_\perp^2|p_{cvz}|^2\over (\omega_\xi n/c)
\sqrt{(\omega_\xi n/c)^2-K^2_\perp}}\right].
\end{eqnarray}
Eq. (82) can be compared with the result from Ref.~\onlinecite{14}
regarding to the exciton energy level in a QW at $H=0$. It turns
out that the dependencies from the components $p_{cvx}, p_{cvy}$
and $p_{cvz}$ coincide and the result $\gamma_{r\xi}=0$ at
$K_\perp>\omega_\xi n/c$ coincides also. The RHS of Eq. (82)
contains the factors $\Omega_0$ and $B_{n_e, n_h}(K_\perp)$ due to
magnetic field. The factor $\left|R_{l_e, l_h}(\sqrt{(\omega_\xi
n/c)^2-K^2_\perp})\right|^2$ is absent in Ref.~\onlinecite{14}
because the QWs with $d<<c/n\omega_l$  have been considered. We
did not apply such a restriction obtaining Eq. (82).

        For the narrow QWs with $d<<\lambda$ one has
\begin{eqnarray}
\label{83}
R_{l_e, l_h}\left(\sqrt{(\omega_\xi n/c)^2-K^2_\perp}
\right)\simeq\nonumber\\
\simeq\pi_{l_e, l_h}=\int_{-\infty}^\infty dz
\varphi_{cle}(z)\varphi_{vlh}(z),
\end{eqnarray}
and for infinitely deep QWs
\begin{equation}
\label{84}
\pi_{l_e, l_h}=\delta_{l_e, l_h}.
\end{equation}
For QWs with $d\simeq c/n\omega_l$
the quantity $\left|R_{l_e, l_h}(\sqrt{(\omega_\xi
n/c)^2-K^2_\perp})\right|^2$ depends on $d$ and diminishes with growth
$d$, therefore the quantity  $\gamma_{r\xi}$ diminishes also.
In the limit $d>>\lambda$ the quantity $\gamma_{r\xi}\rightarrow 0$
as in a bulk crystal.~\cite{12}

        We are interested in obtaining the quantities Eq. (82) at $K_\perp=0$
because we calculate  the absorption and reflection coefficients
for the light normal incidence.
Because $B_{n_e, n_h}(0)=\delta_{n_e, n_h}$,
let us calculate  $\gamma_{r\xi_0}$   for the index set
\begin{equation}
\label{85}
\xi_0\rightarrow c, v, n_e=n_h,  {\bf K}_\perp=0, l_e, l_h.
\end{equation}
With the help of Eq. (82) one obtains
\begin{equation}
\label{86}
\gamma_{r\xi_0}={2e^2\Omega_0\over\hbar cn}
\left|R_{l_e, l_h}(\omega_{\xi_0} n/c)\right|^2
{|p_{cvx}|^2+|p_{cvy}|^2\over m_0\hbar\omega_{\xi_0}},
\end{equation}
where
\begin{equation}
\label{87}
\hbar\omega_{\xi_0}=E_g+\varepsilon^c_{le}+\varepsilon^h_{lh}
+\hbar\Omega_\mu(n_e+1/2),\nonumber\\
\end{equation}
$$\Omega_\mu={|e|H\over\mu c},~~~~~~\mu={m_em_h\over M}.$$

        It follows from Eq. (86) that the EHP inverse radiative lifetime
{\it is proportional to magnetic field} $H$ if the energy
$\hbar\omega_{\xi_0}$ weakly depends on $H$ (approximately
$E_{\xi_0}\simeq{\tilde E}_g,~~~{\tilde E}_g=E_g+\varepsilon^e_{le}+
\varepsilon_{lh}^h)$.

        Let us calculate $\gamma_{r\xi_0}$ for the band model
of $GaAs$. The conductivity band is twice degenerated (on the
spin) and the index $c$ takes two values : $c=1$ or $c=2$.  The
valence band (for heavy holes) is twice degenerate also: $v=1$ or
$v=2$. Two EHP sorts are possible: First, with the indexes $c=1,
v=1$, second, with indexes $c=2, v=2$. These EHP sorts I and II
differ by values ${\bf p}_{cv}$, which are
\begin{equation}
\label{88}
{\bf p}_{cv}^I={1\over\sqrt 2}p_{cv}({\bf e}_x-i{\bf e}_y),~~
{\bf p}_{cv}^{II}={1\over\sqrt 2}p_{cv}({\bf e}_x+i{\bf e}_y).
\end{equation}
For the proposed model the wave functions $\varphi_{cle}(z)$ and
$\varphi_{vlh}(z)$ do not depend on indexes $c$ and $v$.
When the circular polarizations are used,
every polarization (left or right relatively of the axis $z$)
is toughly linked with the EHP sort (I ore II)
because the EHP -- light interaction is proportional to
 ${\bf e}_i{\bf p}_{cv}$.

For the model Eq. (88) one obtains from Eq. (86) the result for
any sort \footnote{The corresponding formula from
Ref.~\onlinecite{25} coincides with Eq. (89) at $n=1$.}
\begin{equation}
\label{89}
\gamma_{r\xi_0}={2e^2\Omega_0\over\hbar cn}
\left|R_{l_e, l_h}(\omega_\xi n/c)\right|^2
{p_{cv}^2\over m_0\hbar\omega_{\xi_0}}.
\end{equation}
Applying the approximation (83) for the narrow QWs with $d<<\lambda$,
we obtain
\begin{equation}
\label{90}
\gamma_{r\xi_0}\simeq{2e^2\Omega_0\over\hbar cn}
\pi^2_{l_e, l_h}.
{p_{cv}^2\over m_0\hbar\omega_{\xi_0}}.
\end{equation}
Applying the parameters for GaAs from
Ref.~\onlinecite{33} and the approximation of Eq. (84)
and supposing $\hbar\omega_{\xi_0}\simeq E_g,$
we obtain the numerical estimation
\begin{equation}
\label{91}
\hbar\gamma_r\simeq 5.35\cdot 10^{-5}{H\over H_{res}} eV,
\end{equation}
where $H_{res}$ appropriates to the magnetopolaron $A$ (Fig. 1),
i. e. equals to $H_{res}=m_ec\omega_{LO}/|e|$.

\section{Inverse radiative lifetimes of magnetopolaron.}

        Let us consider the polaron $A$ (Fig. 1) in a combination
with the hole characterized by the size-quantization quantum
number $l_h$ and the Landau quantum number $n_h=1$. Close to the
resonance at $\omega_{LO} =\Omega_e$ the EHP energy splits into two
magnetopolaron branches with the inverse radiative lifetimes
$\gamma_{ra}$ and $\gamma_{rb}$, according to designations
of Ref.~\onlinecite{26}.
The index $a(b)$ corresponds to the upper (lower) magnetopolaron term.

        To calculate $\gamma_{ra}$ and $\gamma_{rb}$
we will use Eq. (76), determining the function $F_\xi({\bf r})$
from the RHS of Eq. (76).
At first we consider the EHP wave functions under conditions of the
magnetophonon resonance
\footnote{Here and below the indexes $c$ and $v$ are omitted.}
\begin{equation}
\label{92}
\Psi_\Pi({\bf r}_e, {\bf r}_h)|0\rangle=
\psi_{1, k_{hy}, l_h}^h({\bf r}_h)
\Theta_{p, k_{ey}, l_e}({\bf r}_e)|0\rangle.
\end{equation}
The hole wave function $\psi_{1, k_{hy}, l}^h({\bf r})$ is defined in
Eq. (66). The polaron wave function
$\Theta_{p, k_{y}, l}({\bf r}_e)|0\rangle$
is calculated in Ref.~\onlinecite{26}.  The index set in Eq. (92) is
$$\Pi\rightarrow c, v, p; k_{ey}, k_{hy}, l_e, l_h.$$
The index $p$ takes two values : $a$ or $b$.  The function $|0\rangle$
appropriates to the phonon vacuum . The operator $\Theta_{p, k_{y},
l}({\bf r}_e)$, according to Ref.~\onlinecite{26}, can be written
as
\begin{equation}
\label{93}
\Theta_{p, k_{y}, l}({\bf r})=\Theta^0_{p, k_{y}, l}({\bf r})+
\Theta^1_{p, k_{y}, l}({\bf r}),
\end{equation}
where
\begin{equation}
\label{94}
\Theta^0_{p, k_{y}, l}({\bf r})=Q^{1/2}_{0p}
\psi^e_{1, k_{y}, l}({\bf r}),
\end{equation}
\begin{equation}
\label{95}
\Theta^1_{p, k_{y}, l}({\bf r})={Q^{1/2}_{1p}\over|A|}
\sum_\nu e^{ia_Hq_x(k_y-q_y/2)}U^*(\nu)
\psi_{0, k_y-q_y}^e({\bf r})b^+_\nu.
\end{equation}
In the RHS of Eq. (94) and Eq. (95) $\psi^e_{n, k_{y}, l}({\bf r})$ are
the electron wave functions determined in Eq. (66); $\nu$
is the phonon index set, including the transverse component ${\bf q}_\perp$
; $b^+_\nu$ is the phonon creation operator;
$U(\nu)$ is the quantity proportional to electron-phonon interaction
and determined in Ref.~\onlinecite{26};
$A^2=\sum_\nu|U(\nu)|^2$. The polaron wave function
$\Theta_{p,k_{y}, l}({\bf r})|0\rangle$
and the corresponding energy eigenvalue (measured from the energy level
$\varepsilon^e_{le}$)
\begin{eqnarray}
\label{96}
E_p=\hbar\Omega_e
+{1\over 2}\hbar\omega_{LO}\pm\sqrt{(\lambda/2)^2+A^2},\nonumber\\
\lambda=\hbar(\Omega_e-\omega_{LO})
\end{eqnarray}
have been calculated in Ref.~\onlinecite{26}, where it was assumed
that all the LO phonons, composing the polaron, had the same
non-dispersal frequency $\omega_{LO}$. The coefficients in Eq.
(94) and Eq. (95) are
\begin{eqnarray}
\label{97}
Q_{0p}={1\over 2}\left(1\pm{\lambda\over
\sqrt{\lambda^2+4A^2}}\right),\nonumber\\
Q_{1p}={1\over 2}\left(1\mp{\lambda\over
\sqrt{\lambda^2+4A^2}}\right),
\end{eqnarray}
where the upper (lower) sign refers to the term
$p=a~ (p=b)$.

        It follows from Eq. (93) that the polaron wave function is
represented by the linear combination of two functions of the
electron-phonon system,
one of which corresponds to the electron with the Landau quantum
number $n=1$ and to the phonon vacuum, and another
corresponds to the electron with $n=0$ and the LO phonon.
 The coefficients $Q_{0p}$ and $Q_{1p}$
are the probabilities of finding the system
in these states designated with indexes
0 and 1, respectively.

        In the resonance $\Omega_e=\omega_{LO}$
the coefficients  $Q^{res}_{0p}=Q^{res}_{1p}=1/2$ and the energy
\begin{eqnarray}
\label{98}
E_p^{res}=
{3\over 2}\hbar\omega_{LO}\pm\sqrt{A^2},\nonumber\\
\Delta E=E_a^{res}-E_b^{res}=2\sqrt{A^2}.
\end{eqnarray}
To calculate the inverse radiative lifetimes
$\gamma_{ra}$ and
$\gamma_{rb}$ it is necessary to introduce the eigenfunctions
of the EHP momentum operator $\hat{\bf P}$ .
Analogically to Eq. (70) we introduce the functions
\begin{eqnarray}
\label{99}
\Psi_\eta({\bf r}_e, {\bf r}_h)|0\rangle =
\sqrt{{2\pi a^2_H\over L_xL_y}}\times\nonumber\\
\times\sum_{k_{ey}, k_{hy}}
\delta_{K_y, k_{ey}+k_{hy}}e^{-ia_H^2K_xk_y}
\Psi_\Pi({\bf r}_e, {\bf r}_h)|0\rangle,
\end{eqnarray}
characterized by the index set
\begin{equation}
\label{100}
\eta\rightarrow c, v, p; {\bf K}_\perp, l_e, l_h.
\end{equation}
Determining  $\gamma_{r\eta}$ according to Eq. (76),
 we substitute
\begin{equation}
\label{101}
F_\eta({\bf r})=\langle0|\Psi_\eta
({\bf r}_e={\bf r}_h={\bf r})|0\rangle,
\end{equation}
instead of $F_\xi({\bf r})$
and the energy
\begin{equation}
\label{102}
E_\eta=\hbar\omega_\eta=E_g+\varepsilon^e_{le}+\varepsilon^h_{lh}
+{3\over 2}\hbar\Omega_h+E_p.
\end{equation}
instead of the energy $E_\xi$.
So long as $\langle 0|b_\nu^+|0\rangle=0$
 only the first term  from the RHS of Eq.
(93) contributes into the RHS of Eq. (101).
And for $F_\eta({\bf r})$ we obtain the expression
distinguishing from Eq. (75) at $n_e=n_h=1$ only by the additional factor
$Q_{0p}^{1/2}$.  If the $x$ axis is directed along
${\bf K}_\perp$, the result for the inverse radiative lifetimes
$\gamma_{r\eta}$ is analogical to Eq. (82)
\begin{eqnarray}
\label{103}
\gamma_{r\eta}&=&{2Q_{0p}e^2\Omega_0\over\hbar^2cnm_0\omega_\eta}
B_{11}(K_\perp)\times\nonumber\\ &\times&\left|R_{l_e,
l_h}\left(\sqrt{(\omega_\eta n/c)^2-K^2_\perp}
\right)\right|^2\times\nonumber\\
&\times&\left[{\sqrt{(\omega_\eta
n/c)^2-K^2_\perp}\over\omega_\eta n/c}
|p_{cvx}|^2+\right.\nonumber\\ &+&{\omega_\eta
n/c\over\sqrt{(\omega_\eta n/c)^2-K^2_\perp}}
|p_{cvy}|^2+\nonumber\\ &+&\left.{K_\perp^2\over (\omega_\eta n/c)
\sqrt{(\omega_\eta n/c)^2-K^2_\perp}}|p_{cvz}|^2\right];
~K_\perp<\omega_\eta n/c;\nonumber\\
\gamma_{r\eta}&=&0;~~K_\perp>\omega_\eta n/c.
\end{eqnarray}
For the case ${\bf K}_\perp=0$ we introduce the index set
\begin{equation}
\label{104}
\eta_0\rightarrow c, v, p; {\bf K}_\perp=0, l_e, l_h.
\end{equation}
Then for the quantities $\gamma_{r\eta_0}$
we obtain the formulae distinguishing from
Eq. (86), Eq. (89) and Eq. (90) only by the substitution
the indexes  $\eta$ instead of $\xi_0$
(as long as $\omega_{\eta_0}=\omega_\eta$) and by the factor
$Q_{0p}$.

        Instead of Eq. (90) we obtain
\begin{equation}
\label{105}
\gamma_{r\eta_0}\simeq{2Q_{0p}e^2\over\hbar c}
{\Omega_0p_{cv}^2\over nm_0\hbar\omega_\eta}\pi^2_{l_e, l_h}.
\end{equation}
The substitution $\omega_\eta$ instead of $\omega_{\xi_0}$
is not essential as long as approximately
$\omega_{\xi_0}\simeq\omega_\eta\simeq{\tilde
E}_g/\hbar$.  The tough dependence of the inverse radiative lifetime
from the index $p$ and from the value $\lambda$
(characterizing the deviation from the resonance)
is determined by the factor $Q_{0p}$.
In the resonance $Q_{0p}=1/2$, and the quantities $\gamma_{ra}$ and
$\gamma_{rb}$ are equal and contain the factor 1/2 in comparison with
$\gamma_r$ for the appropriate single level (without phonons)
coming through the crossing point of energy terms (see Fig.1).

\section{Non-radiative lifetimes of magnetopolaron}
        We will not calculate the EHP inverse non-radiative lifetimes
in a QW in a magnetic field far away from the magnetopolaron resonance
as long as it is not clear what processes determine them. One-phonon
transitions are forbidden by the energy conservation law. Perhaps
the main contribution is determined by the two-phonon processes
with acoustic phonons participation.
But in the magnetophonon resonance vicinity
{\it \underline{some contribution into}}
$\gamma_p (p=a, b)$, {\it \underline{appears due to the finite
 lifetimes}} {\it \underline{of LO phonons composing the magnetopolaron}}.
We will calculate this contribution here
and determine the low limit of
$\gamma_p (p=a, b)$ in the resonance vicinity.

        The phonon $\nu$ in a QW is characterized by the inverse
non-radiative lifetime $\gamma_\nu$ due to phonon-phonon interaction,
for instance, by the LO phonon decay into two acoustic phonons.

        The quantity $\gamma_\nu$ may be written as
\begin{equation}
\label{106}
\gamma_\nu={2\pi\over\hbar}\sum_f|\langle f|{\cal H}_{ph-ph}|i\rangle|^2
\delta(E_i-E_f),
\end{equation}
where ${\cal H}_{ph-ph}$ is the phonon-phonon interaction,
$|i\rangle=b^+_\nu|0\rangle$ is the initial state of the phonon
$\nu$ with the energy $E_i=\hbar\omega_{LO},
|f\rangle=a^+_\tau|0\rangle$ is the phonon system final state
with the set of indexes $\tau$ and the energy $E_f=E_{\tau},~~~a^+_\tau$
is the phonon operator corresponding, for instance, to the creation
of two acoustic phonons.

        Taking into account that the component
${\bf q}_\perp$ in a QW is preserved in any transitions,
we can write the matrix element in the RHS of Eq.
(106) as
\begin{equation}
\label{107}
\langle 0|a_\tau{\cal H}_{ph-ph}b^+_\nu|0\rangle=
\delta_{{\bf q}_\perp, {\bf q}'_\perp}V({\bf q}_\perp,  j, \vartheta),
\end{equation}
supposing that the initial state index set в $\nu$  is
\begin{equation}
\label{108}
\nu\rightarrow{\bf q}_\perp,  j,
\end{equation}
and the final state index set $\tau$ is
\begin{equation}
\label{109}
\tau\rightarrow{\bf q}_\perp^\prime, \vartheta,
\end{equation}
where ${\bf q}^{\prime}_\perp$ is the resultant transverse component
of the wave vector in the final state, for example,
the sum of the corresponding wave vectors of two phonons;
$j(\vartheta)$ is the initial (final) state index set.
Substituting Eq. (107) into Eq. (106)
and summarizing on ${\bf q}_\perp$, we obtain
\begin{equation}
\label{110}
\gamma_\nu=
{2\pi\over\hbar}\sum_\vartheta
|V({\bf q}_\perp,  j, \vartheta)|^2
\delta(\hbar\omega_{LO}-E_{{\bf q}_\perp, \vartheta}).
\end{equation}
        Applying Eq. (106) we can now calculate the magnetopolaron
inverse non-radiative lifetime $\gamma_{p, k_y, l}$ due to
phonon-phonon interaction.
The initial state wave function is written as
$\Theta_{p, k_y, l}({\bf
r})|0\rangle$, where $|0\rangle$ is a phonon vacuum
, the operator
$\Theta_{p, k_y,
l}({\bf r})$ is determined in Eqs. (93) --- (95).
The final state wave function is the product of the electron and
phonon wave functions
\begin{equation}
\label{111}
|f\rangle=\psi^e_{n', k_y', l'}({\bf r})a^+_\tau|0\rangle,
\end{equation}
where $\psi^e_{n, k_y, l}({\bf r})$ is
 determined in Eq. (66).
Only the second term from the RHS of Eq. (93), describing
the one LO phonon state,
contributes into the matrix element
$\langle f|{\cal H}_{ph-ph}|i\rangle$.
The first term does not contribute because
$\langle 0|a_\tau|0\rangle=0$.
One obtains
\begin{eqnarray}
\label{112}
\langle f|{\cal H}_{ph-ph}|i\rangle&=&
Q_{1p}^{1/2}\sum_\nu e^{ia_H^2q_x(k_y-q_y/2)}{U^*(\nu)\over|A|}
\times\nonumber\\
&\times&\int d{\bf r}\psi^{e*}_{n', k_y', l'}({\bf r})
\psi^e_{0, k_y-q_y, l}({\bf r})\times\nonumber\\
&\times&\langle 0|a_\tau{\cal H}_{ph-ph}b^+_\nu|0\rangle.
\end{eqnarray}
The initial (final) state is characterized
by the indexes $p, k_y, l$
($n', k_y', l', {\bf q}'_\perp , \vartheta$).
The index set  $\nu$ in Eq. (112) consists of
${\bf q}_\perp$ and $j$.
Taking into account the orthonormalization of the electron wave functions
$\psi_{n, k_y, l}^e$ and Eq. (107) and summarizing in
Eq. (112) on ${\bf q}_\perp$, one obtains
\begin{eqnarray}
\label{113}
\langle f|{\cal H}_{ph-ph}|i\rangle&=&
{Q_{1p}^{1/2}\over|A|}\delta_{n', 0}\delta_{l', l}
\delta_{k_y', k_y-q_y'}e^{ia_H^2q_x'(k_x'+q_y'/2)} \times\nonumber\\
&\times&
\sum_jU^*({\bf q}_\perp', j)V({\bf q}_\perp', j, \vartheta).
\end{eqnarray}
Substituting Eq. (113) into Eq. (106) and summarizing
on indexes $n', k_y', l'$
of the final state we obtain
\begin{eqnarray}
\label{114} \gamma_p&=&{2\pi\over\hbar}{Q_{1p}\over A^2}
\sum_{{\bf q}_\perp,\vartheta}\sum_{j_1, j_2} U^*({\bf q}_\perp,
j_1)U({\bf q}_\perp, j_2) V({\bf q}_\perp, j_1,
\vartheta)\times\nonumber\\ &\times&V^*({\bf q}_\perp, j_2,
\vartheta) \delta(E_p-\hbar\Omega_e/2-E_{{\bf q}_\perp,
\vartheta}).
\end{eqnarray}

        Eq. (114) is applicable for the model
used in Ref.~\onlinecite{26}, where it was assumed that the
confined and interface phonons, having the same non-dispersal
frequency $\omega_{LO}$,
 take part in the polaron creation.
The index set $j$ contains the phonon sort indexes.

        It has been shown earlier ~\cite{23} that interaction with
the confined phonons in quite wide QWs may be approximately
replaced by the Fr\"ohlih interaction with the bulk LO phonons and
that interaction with the interface phonons does not contribute
essentially into polaron energy splitting.

        In the case of interaction with the bulk phonons the index
$j$ coincide with
$q_z$, and Eq. (114) is simplified.
One obtains instead of Eq. (107)
\begin{equation}
\label{115}
\langle 0|a_\tau{\cal H}_{ph-ph}b^+_\nu|0\rangle=
\delta_{{\bf q}, {\bf q}'}W({\bf q}, \varphi).
\end{equation}
It is supposed
that the phonon final state is characterized by the 3D
wave vector
${\bf q}$ and by other indexes
$\varphi$. With the help of Eq. (115)  Eq. (114)
is transformed into
\begin{eqnarray}
\label{116}
\gamma_p&=&{2\pi\over\hbar}{Q_{1p}\over A^2}
\sum_{{\bf q}, \varphi}|U({\bf q})|^2|W({\bf q}, \varphi)|^2\times\nonumber\\
&\times&\delta(E_p-\hbar\Omega_e/2-E_{{\bf q}, \varphi}).
\end{eqnarray}
 Eq. (110) for the inverse LO phonon
lifetime (if Eq. (115) is satisfied) can be written as
\begin{equation}
\label{117}
\gamma_{{\bf q}}={2\pi\over\hbar}
\sum_\varphi|W({\bf q}, \varphi)|^2
\delta(\hbar\omega_{LO}-E_{{\bf q}, \varphi}),
\end{equation}
where $E_{{\bf q}, \varphi}$ is the final state energy.

        Applying  Eq. (96) for the polaron term energy
$E_p$ we obtain instead of Eq. (116)
\begin{eqnarray}
\label{118}
\gamma_p&=&{2\pi\over\hbar}{Q_{1p}\over A^2}
\sum_{{\bf q},\varphi}|U({\bf q})|^2|W({\bf q}, \varphi)|^2\times\nonumber\\
&\times&\delta\left(\lambda/2\pm\sqrt{(\lambda/2)^2+A^2}+
\hbar\omega_{LO}-E_{{\bf q}, \varphi}\right),
\end{eqnarray}
where the upper (lower) sign corresponds to the term $a(b)$.
Comparing
Eq. (117) and Eq. (118) one finds that in the polaron resonance vicinity,
i. e. under condition
\begin{equation}
\label{119}
\left|\lambda/2\pm\sqrt{(\lambda/2)^2+A^2}\right|<<\hbar\omega_{LO},
\end{equation}
the quantity $\gamma_p$ is expressed through $\gamma_{\bf q}$,
i. e.
\begin{equation}
\label{120}
\gamma_p=Q_{1p}{\bar\gamma}_{LO},
\end{equation}
\begin{equation}
\label{121}
{\bar{\gamma}}_{LO}=
{\sum_{{\bf q}}\gamma_{{\bf q}}|U({\bf q})|^2\over
\sum_{{\bf q}}|U({\bf q})|^2}.
\end{equation}
Due to the factor $Q_{1p}$ in Eq. (120) the quantity $\gamma_p$
toughly depends on the parameter
$\lambda=\hbar(\Omega_e-\omega_{LO})$.
In the resonance
\begin{equation}
\label{122}
\gamma_p={1\over 2}{\bar\gamma}_{LO}
\end{equation}
for the both terms $p=a$ and $p=b$. One can suppose
that the result from Eq. (114)
is comparable with Eq. (120).

        Let us consider the results of Eq. (118) far away
from the resonance $\Omega_e=\omega_{LO}$.
It is seen on Fig. 1 that the term
$a$ at $\lambda>0$ and the term $b$ at $\lambda<0$ transits
under condition $|\lambda|>>|A|$
into the electron level with indexes $n=1, l$.
According to Eq. (97) the factor $Q_{1p}$
approximately equals to $Q_{1p}\simeq A^2/\lambda^2$.
With the help of
Eq. (118) one obtains approximately
\begin{equation}
\label{123}
\gamma_p={2\pi\over\hbar}{A^2\over\lambda^2}
\sum_{{\bf q},\varphi}{|U({\bf q})|^2\over A^2}
|W({\bf q}, \varphi)|^2
\delta(
\hbar\Omega_e-E_{{\bf q}, \varphi}).
\end{equation}
This value is not expressed through ${\bar\gamma}_{LO}$,
but one can see that in comparison with
${\bar\gamma}_{LO}$ it contains the small factor
$A^2/\lambda^2$. Note that the argument of the
$\delta$-function in Eq. (123) appropriates to the transition
$n=1\rightarrow n=0$ with emitting, for instance,
of two acoustic phonons.
The quantity Eq. (123) is of the second order on phonon-phonon interaction.
It is one of the contributions into the value of the EHP inverse
non-radiative lifetime far away from the magnetophonon resonance.
Two other branches of the terms,
$a$ at $\lambda<0$ and $b$ at $\lambda>0, |\lambda|>>|A|$ ,
appropriate
to the state with the electron on the level $n=0, l$ plus one LO
phonon. $Q_{1p}\simeq 1$ on these branches,
thus we obtain approximately
\begin{equation}
\label{124}
\gamma_p\simeq{\bar\gamma}_{LO},
\end{equation}
as it must be for a state including LO phonon.
Thus the formula
Eq. (118) gives the right limit transitions at $|\lambda|>>|A|$.

        One can conclude that the quantity
$\gamma_p$ increases sharply around the electronic terms intersections
reaching the half of the
LO phonon inverse lifetime.
{\it In the resonance
$\Omega_e=\omega_{LO}$ the value of the inverse
non-radiative lifetime for each of two terms is no smaller than
${\bar\gamma}_{LO}/2$.}

\section{Numerical calculation results}

        The reflection ${\cal R}(\omega_l)$
and absorption ${\cal A}(\omega_l)$ coefficients for the three-level
system and for the different interrelations between
$\gamma_{r1(2)}$, $\gamma_{1(2)}$ and $\omega_1-\omega_2$
are represented in Figs. 2 - 4.
These results have to be used in the case of any two excited levels in a QW
when they are situated quite close each other. When
$\gamma_{r1(2)}<<\omega_1-\omega_2,~~~~\gamma_{1(2)}<<\omega_1-\omega_2$,
the results for the two-level system are applicable.

        Fig. 2 show the dependencies
at $\gamma_{r1}=\gamma_{r2}=\gamma_{r}$,~~~~
$\gamma_1=\gamma_2=\gamma$,~~~~$\gamma_r<<\gamma$.
The curves {\it 1} are relevant to the case
$\gamma<\omega_1-\omega_2$, the curves {\it 2} are relevant to the case
 $\gamma=\omega_1-\omega_2$ and
the curves {\it 3} are relevant to the case
$\gamma>\omega_1-\omega_2$. It is seen that in case 1 the transit
stems to the results for  two-level systems. The maxima are
situated in the vicinity of the points $\omega_l=\omega_1$ and
$\omega_l=\omega_2$.

        In Fig. 3 the functions
${\cal A}(\omega_l)$ and ${\cal R}(\omega_l)$
 for the three level systems
at $\gamma_{r1}=\gamma_{r2}=\gamma_{r}$,~~
$\gamma_1=\gamma_2=\gamma$,~~$\gamma_r>>\gamma$~~are represented.
The curves {\it 1} are relevant to the case
$\gamma_r<\omega_1-\omega_2$,
the curves {\it 2} are relevant to the case
$\gamma_r=\omega_1-\omega_2$
and the curves {\it 3-6} in Fig. 3a
and the curves  {\it 3} in Fig.  3b
are relevant to the case $\gamma_r>\omega_1-\omega_2$.
It is seen that in the case 1 the transit stems to the results
for  two two-level systems.
In other cases the peculiar results can be seen.
In particular the quantity ${\cal R}(\omega_0)$ approaches 0
on all the curves of Fig. 3b in the point
$\omega_l=\omega_0$ .

The reflection and absorption coefficients
are much smaller than one and ${\cal R}(\omega_l)<<{\cal A}(\omega_l)$
as in the case of the two-level systems at
$\gamma_{r1(2)}<<\gamma_{1(2)}$.
On the contrary at $\gamma_{r1(2)}>>\gamma_{1(2)}$
${\cal R}(\omega_l)>>{\cal A}(\omega_l)$
and ${\cal R}$ reaches one.

        In Fig. 4 the functions
${\cal A}(\omega_l)$
and ${\cal R}(\omega_l)$ for the three-level systems
at $\gamma_{r1}=\gamma_{r2}=\gamma_{r}$,~~
$\gamma_1=\gamma_2=\gamma$,~~$\gamma_r=\gamma$~~are represented.
The curves {\it 1} are relevant to the case
$\gamma_r<\omega_1-\omega_2$,
the curves {\it 2} are relevant to the case
$\gamma_r=\omega_1-\omega_2$
and the curves {\it 3} are relevant to the case
$\gamma_r>\omega_1-\omega_2$.
It is seen that the absorption coefficient reaches the maximum values
but does not exceed 0.5.

        In Fig. 5 the energy terms $a$ and $b$ for the system
consisting of the polaron $A$ (Fig. 1) and the hole with
indexes $n=1, l$ are represented.
$\Delta E$ is term's splitting at $\lambda=0$.
${\cal E}_p$ is the system energy
measured from
$$E_0=E_g+\varepsilon_l^e+\varepsilon_l^h+(3/2)(1+m_e/m_h)
\hbar\omega_{LO}.$$
According to  Eq. (102) and Eq. (96),
\begin{equation}
\label{125}
{{\cal E}_p\over\Delta E}={\lambda\over\Delta E}
\left(1+{3\over 2}{m_e\over m_h}\right)\pm{1\over 2}\sqrt{1+
\left({\lambda\over\Delta E}\right)^2},
\end{equation}
where the upper (lower) sign
corresponds to the term $a$($b$).
The value $m_e/m_h=0.2$ is applied, which is relevant to GaAs.
 Calculating
${\cal A}(\omega_l)$ and ${\cal R}(\omega_l)$
for magnetopolarons in QWs
one has to substitute
the quantities ${\cal E}_a$ and ${\cal E}_b$
(see Eq. (125)) into all the formulae of the section
IY as energies $\hbar\omega_1$ and $\hbar\omega_2$
when
$\hbar\omega_l$  is measured from the level $E_0$.

        In Fig. 6 the dependencies
$\gamma_{rp}$ and $\gamma_p/\gamma_{rp}$ from the magnetic field
value are represented. Fig. 9 (10) appropriates to the term $a$
($b$). The constant $\gamma_0$ is the EHP inverse radiative
lifetime at $\lambda=0$ without the polaron effect. According to
the section Y,
\begin{equation}
\label{126} \gamma_0={2e^2\omega_{LO}m_e\over\hbar
cnm_0}{p^2_{cv}\over m_0{\tilde{E}}_g}\pi^2_{ll}.
\end{equation}
With the help of Eq. (105) we obtain
\begin{equation}
\label{127} {\gamma_{rp}\over\gamma_0}={1\over
2}\left(1+{\lambda/\Delta E\over \Delta E/\hbar\omega_{LO}}\right)
\left[1\pm{\lambda/\Delta E\over\sqrt{1+(\lambda/\Delta
E)^2}}\right],
\end{equation}
which was applied in Fig. 6.
The upper (lower) sign is relevant to the term $p=a$ ($b$).
The factor in the parenthesis in the RHS of Eq.
(127) describes the magnetic field dependence of
$\Omega_0$ entering in $\gamma_{rp}$ (see Eq. (105)).
 This factor depends on
$\Delta E/\hbar\omega_{LO}$,
which is chosen equal to 0.18.
\footnote{According to Ref.~\onlinecite{23} in GaAs (the polaron $A$)
at $d=300 A$
energy splitting is $\Delta E\simeq 6.65 10^{-3} eV$.
As long as in GaAs  at
$\hbar\omega_{LO}=0.0367 eV$ one obtains $\Delta E/\hbar\omega_{LO}
\simeq 0.181.$}

        According to Eq. (105) and Eq. (120)
\begin{equation}
\label{128} {\gamma_p\over
\gamma_{rp}}={{\bar{\gamma}}_{LO}/\gamma_0\over 1+{\lambda/\Delta
E\over\Delta E/\hbar\omega_{LO}}} {1\mp{\lambda/\Delta
E\over\sqrt{1+(\lambda/\Delta E)^2}} \over 1\pm{\lambda/\Delta
E\over\sqrt{1+(\lambda/\Delta E)^2}}},
\end{equation}
where the upper (lower) sign appropriates to the term $p=a$ ($b$).

        Eq. (127) and  Eq. (128) were applied in Figs. 9, 10.
On Figs. 9, 10 the function $\gamma_p/\gamma_{rp}$ for the three
values of $\bar\gamma_{LO}/\gamma_0$ is represented: 10 (curve
{\it 1}), 1 (curve {\it 2}), 0.1 (curve {\it 3}). According to our
estimate with applying the parameter values from
Ref.~\onlinecite{32}
\begin{equation}
\label{129} \hbar\gamma_0\simeq 5.35\cdot 10^{-5} eV.
\end{equation}
Eqs. (125) - (128) and Fig. 5 - 6
allow us in principle to determine
${\cal A}(\omega_l)$ and ${\cal R}(\omega_l)$ with the help
of the results of the section IY at a fixed value of magnetic field.
One can also determine ${\cal A}(H)$ and ${\cal R}(H)$
 at fixed $\omega_l$ values.

The quantity $\bar{\gamma}_{LO}$ remains practically unknown.
\footnote{One can try to estimate $\bar{\gamma}_{LO}$
on the line width of one-phonon scattering in the bulk GaAs.}
Therefore on Figs. 6 - 8 the results for the different interrelations
$\bar{\gamma}_{LO}/\gamma_0$ are represented.

        One can conclude from Fig. 5 that the
magnetopolaron effect area spreads approximately
from $\lambda/\Delta E=-2$ to $\lambda/\Delta E=2$.
At $\lambda/\Delta E>2$ the term $a$ (and at
$\lambda/\Delta E<-2$ the term $b$)
transits into the EHP
energy level with quantum numbers of the electron and hole
$n_e=n_h=1,~~~l_e=l_h=l.$ The term $b$ at $\lambda/\Delta E>2$ and
the term $a$  at
$\lambda/\Delta E<-2$ transit into the energy level
of the system consisting of the EHP with quantum numbers
$n_e=0, n_h=1,~~~l_e=l_h=l$ and one LO phonon.
This state interacts weakly with exciting light out of the
magnetophonon resonance.

        In the case of the polaron $A$ the value $\Delta E$ of
energy splitting is very large. Therefore the conditions
$\gamma_{rp}<<\omega_1-\omega_2,~~~\gamma_p<<\omega_1-\omega_2$
are obviously satisfied and the approximation of two
two-level systems is applicable for calculations of
${\cal A}$ and ${\cal R}$, i. e. Eq. (36) and Eq. (37)
of the section IY.  An application of the results of the same
section, relevant to the three-level systems, is necessary in the
case of smaller values $\Delta E$, which would observed, for instance,
in the case of weak polarons (see the section II).

        In the magnetopolaron area, two maxima have to be observed
in the functions ${\cal A}(\omega_l)$ and ${\cal R}(\omega_l)$ at
the definite value  $H$ of the magnetic field.  The maximum points
$\omega_{la}$ and $\omega_{lb}$ one can determine
, drawing the vertical sections on Fig. 5.
Two maxima have to be observed also in functions
  ${\cal A}(H)$ and ${\cal R}(H)$ at $\omega_l=const$.
The appropriate magnetic field values
$H_a$ and $H_b$ can be determined
drawing the horizontal sections on Fig. 5.
       The quantities $\gamma_{rp}$ and $\gamma_p$
are strongly dependent on magnetic field in the magnetophonon resonance
vicinity (see Fig. 6).

        The results for
${\cal A}(\omega_l)$ and ${\cal R}(\omega_l)$ are different for
the different values of ${\bar{\gamma}}_{LO}/\gamma_0$. At those
values of $\lambda/\Delta E$ (i.e. the magnetic fields $H$), where
$\gamma_p/\gamma_{rp}>>1$, light absorption is much smaller than
one , but exceeds light reflection.  In the case of curves {\it 1}
on Figs. 9,10, relevant to the parameter
${\bar{\gamma}}_{LO}/\gamma_0=10$, this area for the both terms
$a$ and $b$ takes almost the whole interval of magnetic field
values where the polaron effect is essential. Indeed, for the
curves {\it 1} in Fig. 6~~~ $\gamma_a/\gamma_{ra}=1$ at
$\lambda_{a1}/\Delta E=1.24$~~~, and $\gamma_b/\gamma_{rb}=1$ at
$\lambda_{b1}/\Delta E=-1.82$~~~.  Around the points
$\lambda_{a1}$
 and $\lambda_{b1}$ light reflection by terms $a$ and $b$
reaches the largest values (${\cal A}$=1/2 and ${\cal R}$=1/4 in
maximum). In the regions $\lambda>>\lambda_{a1}$, and
$\lambda<<\lambda_{b1}$ the polaron effect is inessential. The
second maxima on the curves $A(\omega_l)$ and $R(\omega_l)$
disappear. These results are alike the results for the EHP with
$n_e=n_h=1,~~~l_e=l_h=l$ far from the polaron resonance. In this
area $\gamma_p/\gamma_{rp}<<1$, and light reflection exceeds light
absorption.

        In the opposite case ${\bar{\gamma}}_{LO}/\gamma_0=0.
1$ (curves {\it 3}) the condition $\gamma_p/\gamma_{rp}<<1$ is
satisfied in almost the whole region of the magnetophonon
resonance.

        This region includes the resonant point
$\lambda=0$.  Light absorption is much smaller than light
reflection which reaches the values ${\cal R}=1$ in both maxima on
the curve ${\cal R}(\omega_l)$. At $\lambda_{a3}/\Delta
E=-1.24$~~~ and $\lambda_{b3}/\Delta E=1.68$~~~ the inverse
non-radiative and radiative lifetimes become equal , which
corresponds to maximum absorption. On the left of the point
$\lambda_{a3}$ the term $a$ does not interact with light (note the
small values $\gamma_{ra}$).  The same is relevant to the term $b$
on the right of $\lambda_{b3}$.

        The curves {\it 2} on Figs. 6 are relevant
to an intermediate case ${\bar{\gamma}}_{LO}/\gamma_0=1.$ The
maximum absorption point coincides with the resonance point
$\lambda=0$. For the term $a$ at $\lambda>0$    the condition
$\gamma_a/\gamma_{ra}<1$ is satisfied, at $\lambda<0$  the
opposite condition $\gamma_a/\gamma_{ra}>1$ is satisfied. There is
the opposite picture for the term $b$. At $\gamma_p/\gamma_{rp}<1$
light reflection dominates, at $\gamma_p/\gamma_{rp}>1$
 light absorption dominates .

        Figs. 7, 8 ${\cal A}(H)$ and ${\cal R}(H)$ in
the polaron splitting vicinity. The curves {\it 1, 2, 3}
correspond to three different magnitudes of the frequency
$\omega_l$.
The Figs. 7a and 8a correspond to the ratio
${\bar\gamma}_{LO}/\gamma_0=10$.
The Figs. 7b and 8b correspond to the ratio
${\bar\gamma}_{LO}/\gamma_0=0.4$.

        To summarize, a classification of magnetopolarons in a QW
in strong magnetic field has been done. The usual polarons
(including double-, triple- and so on), combined polarons (for
which the resonant magnetic field depends on the QW width and
depth) as well as the weak polarons (for which a polaron energy
splitting $\Delta E$ is comparatively small) have been defined.

        The formulae for the non-dimensional coefficients of light
absorption and reflection by QW under normal incidence of light have
been obtained for the three-level electronic systems. Our
calculations are based on taking into account the radiative
lifetimes of the electronic systems on two excited level, which
means taking into account all of the absorption and reradiation
processes and a transcendence of the perturbation theory on the
light-electron coupling constant.

        The EHP radiative lifetime in a QW in a strong magnetic field
under an arbitrary value of the QW inplane EHP wave vector ${\bf
K}_\perp$ has been calculated far from the magnetophonon
resonance. The radiative lifetimes  of a system, consisting of a
magnetopolaron and a hole, have been calculated.

     The expressions for the contributions into non-radiative
lifetimes of the polaron states due to the finite LO phonon
lifetimes are obtained. It turned out that the radiative lifetimes
of polarons as well as non-radiative ones are strongly dependent on
magnetic field $H$ in the magnetopolaron resonance vicinity.

        The functions  ${\cal A}(H)$ and ${\cal R}(H)$ have been obtained
by taking into account the magnetopolaron effect.

\section{Acknowledgements}
        S.T.P thanks the Zacatecas University and the National
Council of Science and Technology (CONACYT) of Mexico for the financial
support and hospitality. D.A.C.S. thanks CONACYT (27736-E) for the financial
support.
       Authors are grateful to V. M. Ba\~nuelos-Alvarez for the computer
cooperation and to A. D'Amor for a critical reading of the manuscript.
       This work has been partially supported by the Russian
Foundation for Basic Research and by the Program "Solid State
Nanostructures Physics".


\newpage

\begin{figure}

\caption{The energy levels of an electron (hole)--LO phonon system
as a function of the cyclotron frequency $\Omega$.
$E$  is the energy measured from the size-quantization
energy level $\varepsilon$. The double,  triple,
quaternate, weak and combined polarons
are marked with the filled circles, triangles,
squares, empty circles and crossed circles, respectively.}

\end{figure}

\begin{figure}

\caption{Non-dimensional light absorption
${\cal A}$ and reflection ${\cal R}$ coefficients
of a three-level system  as a function
of the light frequency $\omega_l$ for the case
$\gamma>>\gamma_r$, where  $\gamma=\gamma_1=\gamma_2$
is the non-radiative homogeneous level broadening,
$\gamma_r=\gamma_{r1}=\gamma_{r2}$ is the radiative level
broadening. Parameter magnitudes have been used:
 $\hbar(\omega_1
-\omega_2)=0.005,~~~\hbar\gamma_r=0.0001,~~~\hbar\gamma=0.002$
(curve {\it 1}),
$\hbar\gamma=0.005$ (curve {\it 2}) и $\hbar\gamma=0.008$ (curve {\it 3})
in eV or in arbitrary units.}

\end{figure}

\begin{figure}

\caption{Same as Fig. 2 for the case $\gamma<<\gamma_r.$
The parameter magnitudes have been used:
 $\hbar(\omega_1-\omega_2)=0.005,~~~
\hbar\gamma=0.0001.$
Fig. 3a:
$\hbar\gamma_r=0.002$ (curve {\it 1}) ,
 $~~~\hbar\gamma_r=0.005$ (curve {\it 2})
and  $~~\hbar\gamma_r=0.008$ (curve {\it 3}).
On the inset: $\hbar\gamma_r=0.04$ (curve {\it 4}),
  $\hbar\gamma_r=0.125$ (curve {\it 5}) and
 $\hbar\gamma_r=0.5$ (curve {\it 6}).
Fig. 3b: $\hbar\gamma_r=0.002$ (curve {\it 1}),
  $\hbar\gamma_r=0.005$ (curve {\it 2}) and
 $\hbar\gamma_r=0.008$ (curve {\it 3}).}

\end{figure}

\begin{figure}

\caption{Same as Fig. 2 for the case $\gamma=\gamma_r.$
The absorption (a) and reflection (b) coefficients.
The parameter magnitudes have been  used:
 $\hbar(\omega_1-\omega_2)=0.005,~~~
\hbar\gamma=\hbar\gamma_r=0.002$ (curve {\it 1});
$\hbar\gamma=\hbar\gamma_r=0.005$
(curve {\it 2}), $\hbar\gamma=\hbar\gamma_r=0.008$ (curve {\it 3}).}
\end{figure}

\begin{figure}

\caption{Excitation energies of an electron system consisting of a
type-$A$ polaron  (see Fig. 1) and a hole with quantum numbers
$n_h=1,~~l$ as a function of magnetic field. The parameter
$m_e/m_h=0.2$ (corresponding to GaAs) has been used. The energy
${\cal E}_p$ is measured from the level
$E_0=E_g+\varepsilon^e_l+\varepsilon^h_l+(3/2)(1+m_e/m_h)
\hbar\omega_{LO}, \Delta E$ is the polaron energy splitting in the
point $\Omega_e=\omega_{LO}.$}

\end{figure}

\begin{figure}

\caption{The radiative and non-radiative inverse lifetimes of
electronic excitations in a QW consisting of a polaron $A$ and a
hole with quantum numbers $n_h=1$ and $l$ as a function of
magnetic field. The parameter $\Delta E/\hbar\omega_{LO}=0.18$
(corresponding to GaAs, Ref. [23]) has been used.
$\gamma_{rp}/\gamma_0$ (solid lines),
 $\gamma_p/\gamma_{rp}$ (dashed lines).
The index $p$ designates $a$ and $b$.
Fig. 6a(b) appropriates to  $p=a(b)$.
The parameter magnitudes have been  used for the dashed lines:
$\bar{\gamma}_{LO}/\gamma_0=10$ (curve {\it 1}),
$\bar{\gamma}_{LO}/\gamma_0=1$ (curve {\it 2}) and
$\bar{\gamma}_{LO}/\gamma_0 =0.1$ (curve {\it 3}).}

\end{figure}

\begin{figure}

\caption{The light absorption coefficient ${\cal A}$ in a polaron
splitting vicinity as a function of magnetic field. The polaron
$A$ (see Fig. 1) parameters for GaAs have been used: $\Delta
E=6.65\cdot 10^{-3}$ eV (Ref. 23), $\gamma_0=10^{-4} eV.$ $a$
corresponds to $\bar{\gamma}_{LO}/\gamma_0=10$, $b$ corresponds to
$\bar{\gamma}_{LO}/\gamma_0=0.4$. The curves {\it 1}, {\it 2} and
{\it 3} distinguish with different values of $\omega_l$.
$\Delta\omega_l=\omega_l-(\varepsilon_a + \varepsilon_b)/2\hbar=0$
(curves {\it 1}),  $\Delta\omega_l= \Delta E/2$ (curves {\it 2})
and $\Delta\omega_l=-\Delta E/2$ (curves {\it 3}).}

\end{figure}

\begin{figure}

\caption{The light reflection coefficient {\cal R} in the polaron
splitting vicinity as a function of magnetic field.
Parameters and designations of Fig. 7 have been used.}

\end{figure}
\end{document}